\documentclass[aps,apl,twocolumn,groupedaddress,amsmath]{revtex4-1}
\usepackage[latin1]{inputenc}    % Accept european-encoded (latin1) characters.
\usepackage{graphicx}   % For eps figures
\usepackage{xspace}
\usepackage{color}

\begin{document}

% some macros
%-----------------------------------------------
%\def\Tone{$\mathrm{T_1}$}
%\def\Ttwo{$\mathrm{T_2}$}
%\def\TtwoE{$\mathrm{T^E_2}$}
%\def\Tph{$\mathrm{T^E_{\phi}}$}
%\newcommand{\mpar}[1]{\marginpar{\small \textcolor{magenta}{#1}}}
\newcommand{\mpar}[1]{\marginpar{\small \it \textcolor{magenta}{#1}}}

\newcommand*{\siX}{\ensuremath{\sigma_\mathrm{x}}\xspace}
\newcommand*{\siZ}{\ensuremath{\sigma_\mathrm{z}}\xspace}

\newcommand*{\Isq}{\ensuremath{I_\mathrm{b}}\xspace}
\newcommand*{\PhiQ}{\ensuremath{\Phi_\mathrm{qb}}\xspace}
\newcommand*{\fqb}{\ensuremath{f_\mathrm{qb}}\xspace}
\newcommand*{\fr}{\ensuremath{f_\mathrm{r}}\xspace}
\newcommand*{\tp}{\ensuremath{t_\mathrm{p}}\xspace}
\newcommand*{\fRabi}{\ensuremath{f_\mathrm{Rabi}}\xspace}
\newcommand*{\Ian}{\ensuremath{I_\mathrm{antenna}^\mathrm{mw}}\xspace}
\newcommand*{\Imr}{\ensuremath{I_\mathrm{r}^\mathrm{mw}}\xspace}
\newcommand*{\emd}{\ensuremath{\varepsilon^\mathrm{mw}_\mathrm{direct}}\xspace}
\newcommand*{\emt}{\ensuremath{\varepsilon^\mathrm{mw}}\xspace}

\newcommand*{\PhiX}{\ensuremath{\Phi_\mathrm{X}}\xspace}
\newcommand*{\PhiZ}{\ensuremath{\Phi_\mathrm{Z}}\xspace}
\newcommand*{\fX}{\ensuremath{f_\mathrm{x}}\xspace}
\newcommand*{\fZ}{\ensuremath{f_\mathrm{z}}\xspace}
\newcommand*{\Ax}{\ensuremath{A_\mathrm{x}}\xspace}
\newcommand*{\Az}{\ensuremath{A_\mathrm{z}}\xspace}

\newcommand*{\TF}{\ensuremath{T_{\varphi F}}\xspace}
\newcommand*{\TE}{\ensuremath{T_{\varphi E}}\xspace}
\newcommand*{\GF}{\ensuremath{\Gamma_{\varphi F}}\xspace}
\newcommand*{\GE}{\ensuremath{\Gamma_{\varphi E}}\xspace}

\newcommand*{\um}{\ensuremath{\,\mu\mathrm{m}}\xspace}
\newcommand*{\nm}{\ensuremath{\,\mathrm{nm}}\xspace}
\newcommand*{\mm}{\ensuremath{\,\mathrm{mm}}\xspace}
\newcommand*{\m}{\ensuremath{\,\mathrm{m}}\xspace}
\newcommand*{\sqm}{\ensuremath{\,\mathrm{m}^2}\xspace}
\newcommand*{\sqmm}{\ensuremath{\,\mathrm{mm}^2}\xspace}
\newcommand*{\squm}{\ensuremath{\,\mu\mathrm{m}^2}\xspace}
\newcommand*{\psqm}{\ensuremath{\,\mathrm{m}^{-2}}\xspace}
\newcommand*{\psqmV}{\ensuremath{\,\mathrm{m}^{-2}\mathrm{V}^{-1}}\xspace}
\newcommand*{\cm}{\ensuremath{\,\mathrm{cm}}\xspace}

\newcommand*{\nF}{\ensuremath{\,\mathrm{nF}}\xspace}
\newcommand*{\pF}{\ensuremath{\,\mathrm{pF}}\xspace}
\newcommand*{\pH}{\ensuremath{\,\mathrm{pH}}\xspace}

\newcommand*{\emob}{\ensuremath{\,\mathrm{m}^2/\mathrm{V}\mathrm{s}}\xspace}
\newcommand*{\edos}{\ensuremath{\,\mu\mathrm{C}/\mathrm{cm}^2}\xspace}
\newcommand*{\mbar}{\ensuremath{\,\mathrm{mbar}}\xspace}

\newcommand*{\A}{\ensuremath{\,\mathrm{A}}\xspace}
\newcommand*{\mA}{\ensuremath{\,\mathrm{mA}}\xspace}
\newcommand*{\nA}{\ensuremath{\,\mathrm{nA}}\xspace}
\newcommand*{\pA}{\ensuremath{\,\mathrm{pA}}\xspace}
\newcommand*{\fA}{\ensuremath{\,\mathrm{fA}}\xspace}
\newcommand*{\uA}{\ensuremath{\,\mu\mathrm{A}}\xspace}

\newcommand*{\Ohm}{\ensuremath{\,\Omega}\xspace}
\newcommand*{\kOhm}{\ensuremath{\,\mathrm{k}\Omega}\xspace}
\newcommand*{\MOhm}{\ensuremath{\,\mathrm{M}\Omega}\xspace}
\newcommand*{\GOhm}{\ensuremath{\,\mathrm{G}\Omega}\xspace}

\newcommand*{\Hz}{\ensuremath{\,\mathrm{Hz}}\xspace}
\newcommand*{\kHz}{\ensuremath{\,\mathrm{kHz}}\xspace}
\newcommand*{\MHz}{\ensuremath{\,\mathrm{MHz}}\xspace}
\newcommand*{\GHz}{\ensuremath{\,\mathrm{GHz}}\xspace}
\newcommand*{\THz}{\ensuremath{\,\mathrm{THz}}\xspace}

\newcommand*{\K}{\ensuremath{\,\mathrm{K}}\xspace}
\newcommand*{\mK}{\ensuremath{\,\mathrm{mK}}\xspace}

\newcommand*{\kV}{\ensuremath{\,\mathrm{kV}}\xspace}
\newcommand*{\V}{\ensuremath{\,\mathrm{V}}\xspace}
\newcommand*{\mV}{\ensuremath{\,\mathrm{mV}}\xspace}
\newcommand*{\uV}{\ensuremath{\,\mu\mathrm{V}}\xspace}
\newcommand*{\nV}{\ensuremath{\,\mathrm{nV}}\xspace}

\newcommand*{\eV}{\ensuremath{\,\mathrm{eV}}\xspace}
\newcommand*{\meV}{\ensuremath{\,\mathrm{meV}}\xspace}
\newcommand*{\ueV}{\ensuremath{\,\mu\mathrm{eV}}\xspace}

\newcommand*{\T}{\ensuremath{\,\mathrm{T}}\xspace}
\newcommand*{\mT}{\ensuremath{\,\mathrm{mT}}\xspace}
\newcommand*{\uT}{\ensuremath{\,\mu\mathrm{T}}\xspace}

\newcommand*{\ms}{\ensuremath{\,\mathrm{ms}}\xspace}
\newcommand*{\s}{\ensuremath{\,\mathrm{s}}\xspace}
\newcommand*{\us}{\ensuremath{\,\mathrm{\mu s}}\xspace}
\newcommand*{\ns}{\ensuremath{\,\mathrm{ns}}\xspace}
\newcommand*{\rpm}{\ensuremath{\,\mathrm{rpm}}\xspace}
\newcommand*{\minute}{\ensuremath{\,\mathrm{min}}\xspace}
\newcommand*{\degree}{\ensuremath{\,^\circ\mathrm{C}}\xspace}

\newcommand*{\EqRef}[1]{Eq.~(\ref{#1})}
\newcommand*{\FigRef}[1]{Fig.~\ref{#1}}
\newcommand*{\dd}[2]{\mathrm{\partial}#1/\mathrm{\partial}#2}
\newcommand*{\ddf}[2]{\frac{\mathrm{\partial}#1}{\mathrm{\partial}#2}}

\title{Driven dynamics and rotary echo of a qubit tunably coupled to a harmonic oscillator}
 \author{S. Gustavsson$^1$}
% \email{simongus@mit.edu}
 \author{J. Bylander$^1$}
 \author{F. Yan$^2$}
 \author{P. Forn-D\'{\i}az$^{1,\dag}$}
 \author{V. Bolkhovsky$^3$}
 \author{D. Braje$^3$}
 \author{G. Fitch$^3$}
 \author{K. Harrabi$^{4,\bigtriangledown}$}
 \author{D. Lennon$^3$}
 \author{J. Miloshi$^3$}
 \author{P. Murphy$^3$}
 \author{R. Slattery$^3$}
 \author{S. Spector$^3$}
 \author{B. Turek$^{3,*}$}
 \author{T. Weir$^3$}
 \author{P.B. Welander$^3$}
 \author{F. Yoshihara$^4$}
 \author{D.G. Cory$^{5,6}$}
 \author{Y. Nakamura$^{4,7,\ddag}$}
 \author{T.P. Orlando$^{1}$}
 \author{W.D. Oliver$^{1,3}$}
 \affiliation{$^1$Research Laboratory of Electronics, Massachusetts Institute of Technology, Cambridge, MA 02139, USA \\
  $^2$Department of Nuclear Science and Engineering, MIT, Cambridge, MA 02139, USA \\
  $^3$MIT Lincoln Laboratory, 244 Wood Street, Lexington, MA 02420, USA \\
  $^4$The Institute of Physical and Chemical Research (RIKEN), Wako, Saitama 351-0198, Japan \\
  $^5$Institute for Quantum Computing and Department of Chemistry, University of Waterloo, Ontario, Canada\\
  $^6$The Perimeter Institute for Theoretical Physics, Ontario, Canada\\
  $^7$Green Innovation Research Laboratories, NEC Corporation, Tsukuba, Ibaraki 305-8501, Japan\\
  $^\dag$Present address:  Norman Bridge Laboratory of Physics, California Institute of Technology, Pasadena, California 91125, USA\\
  $^\bigtriangledown$Present address:  Physics Department, King Fahd University of Petroleum \& Minerals, Dhahran 31261, Saudi Arabia\\
  $^*$Present address: The Johns Hopkins University Applied Physics Laboratory, 11100 Johns Hopkins Road, Laurel, MD 20723, USA\\
  $^\ddag$Present address: Dept. of Applied Physics, University of Tokyo, Hongo, Bunkyo-ku, Tokyo 113-8656, Japan}

%\date{\today}
\begin{abstract}
We have investigated the driven dynamics of a superconducting flux qubit that is tunably coupled to a microwave resonator.
We find that the qubit experiences an oscillating field mediated by off-resonant driving of the resonator, leading to strong modifications of the qubit Rabi frequency. This opens an additional noise channel, and we find that low-frequency noise in the coupling parameter causes a reduction of the coherence time during driven evolution. The noise can be mitigated with the rotary-echo pulse sequence, which, for driven systems, is analogous to the Hahn-echo sequence.
\end{abstract}

\maketitle
% introduction

Circuit quantum electrodynamics implemented with superconducting artificial atoms and microwave resonators
has emerged as a framework for studying on-chip light-matter interactions \cite{Blais:2004,Wallraff:2004,Chiorescu:2004}.
It has enabled a range of experiments %involving microwave resonators, waveguides, and artificial atoms,
 including lasing \cite{Astafiev:2007}, the creation  \cite{Houck:2007,Hofheinz:2008,Hofheinz:2009} and detection \cite{Schuster:2007} of arbitrary Fock states, and microwave photon-correlation measurements \cite{Bozyigit:2010,Mallet:2011}.
% - importance of resonator for qubit state transfer
Microwave resonators also provide a means to couple distant qubits \cite{Sillanpaa:2007,Majer:2007} and, in this role, have been used to implement quantum algorithms in superconducting circuits \cite{DiCarlo:2009} and to develop quantum computer architectures \cite{Mariantoni:2011}.
However, the coupling of a qubit to a resonator also influences the qubit coherence, for example by modifying its relaxation rate through the Purcell effect \cite{Houck:2008}.

In this work, we study an additional consequence of the resonator by investigating the driven dynamics and the dephasing of a flux qubit \cite{Mooij:1999} that is tunably coupled to a harmonic oscillator \cite{Allman:2010, Fedorov:2010, Srinivasan:2011, Gambetta:2011}. We find that the resonator mediates an indirect driving field that interferes with the direct drive set by the qubit-antenna coupling, thereby modifying both the amplitude and the phase of the net driving field. % for non-zero qubit-resonator coupling,
%
%The effect applies to any type of superconducting qubit coupled to a resonator.
The tunable coupling allows the indirect driving to be switched off, but it also opens an additional channel for noise to couple into the system. Fluctuations in the coupling parameter translate into effective driving-field amplitude noise, which reduces the qubit coherence during driven evolution.
We show that the qubit dephasing due to amplitude noise (whether due to tunable coupling or otherwise) can be mitigated by a rotary echo \cite{Solomon:1959}, a pulse sequence originally developed for nuclear magnetic resonance.

\begin{figure}[t]
\centering
\includegraphics[width=\linewidth]{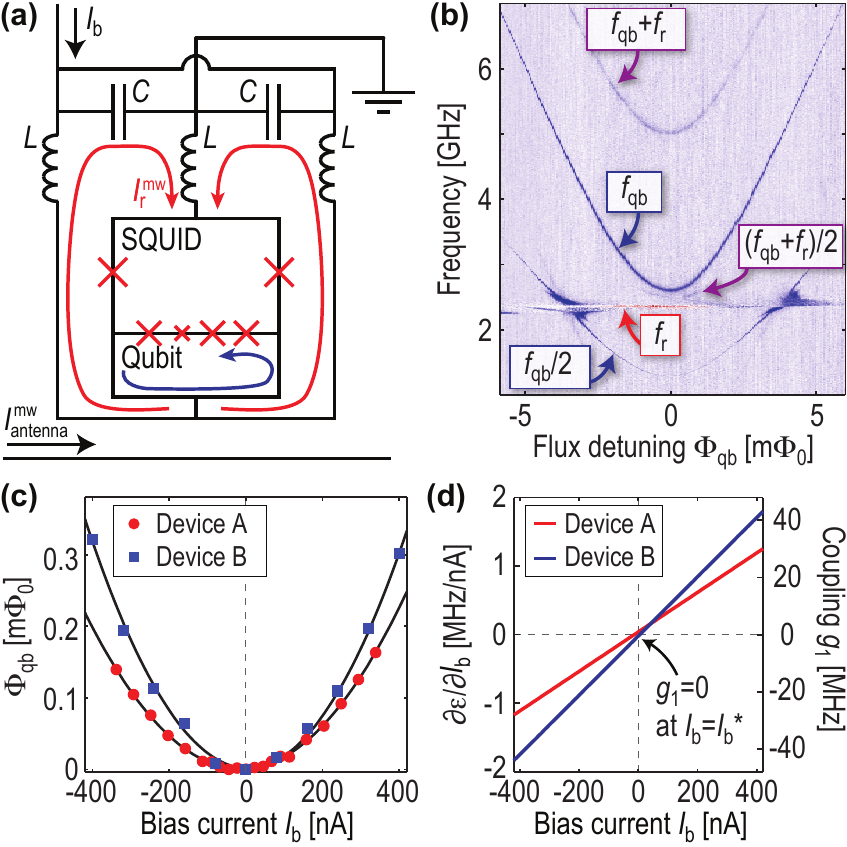}
\caption{(a) Circuit diagram of the qubit and the oscillator.
 The qubit state is encoded in currents circulating clockwise or counterclockwise in
 the qubit loop (blue arrow), while the mode of the harmonic oscillator is shown by the red arrows.
 %Typical sample parameters are $L=130\pH$ and $C = 7\pF$.
 (b) Spectrum for device A, showing the qubit and the harmonic oscillator.
 In addition, the two-photon qubit ($\fqb/2$) and the qubit+resonator ($\fqb+\fr$) transitions are visible.
 (c) Flux induced in the qubit loop by the dc bias current $\Isq$. The black lines are parabolic fits.
  (d) First-order coupling between the qubit and the ground state of the harmonic oscillator, showing that the coupling is tunable by adjusting $\Isq$. The coupling is zero at $\Isq = \Isq^*$, which is slightly offset from $\Isq=0$ due to fabricated junction asymmetry.
 The derivative $\dd{\varepsilon}{\Isq}$ is calculated from the curves in panel (c). The qubit parameters are: $I_\mathrm{P} = 175\nA$ for device A and $I_\mathrm{P} = 180\nA$ for device B.  The resonators have quality factors  $Q \approx 100$.
 The right-hand axis is calculated using $\fr = 2.2\GHz$ and $C_\mathrm{eff} = 2C = 14\pF$ for both samples.
}
\label{fig:Sample}
\end{figure}

% samples
The device, shown in \FigRef{fig:Sample}(a), consists of a flux qubit and a SQUID embedded in a two-mode LC resonant circuit.  The diabatic states of the qubit correspond to clockwise or counterclockwise persistent currents in the qubit loop [blue arrow in \FigRef{fig:Sample}(a)], with energies controlled by the flux in the loop.
The resonator mode of interest is the SQUID plasma mode, depicted by the two red arrows in \FigRef{fig:Sample}(a).
%The resonant circuit has a symmetric design, aiming at minimizing its coupling to the antenna, while keeping the antenna-qubit coupling strong enough to enable fast qubit operations.
%
% - readout
The SQUID serves dual purposes: it acts as a tunable coupler between the qubit and the resonator, and it is also used as a sensitive magnetometer for qubit read-out \cite{Chiorescu:2003}.
%The read-out is implemented by applying a short current pulse to the SQUID and monitoring the voltage response; due to the inductive coupling between the SQUID and the qubit, the SQUID switching probability will vary depending on the qubit state \cite{Chiorescu:2003}.

% spectrum
We have investigated two devices with similar layouts but slightly different parameters, both made of aluminum. Device A was designed and fabricated at MIT Lincoln Laboratory and device B was designed and fabricated at NEC.
Figure \ref{fig:Sample}(b) shows a spectroscopy measurement of device A versus applied flux, with the qubit flux detuning $\PhiQ$ defined as $\PhiQ = \Phi + \Phi_0/2$ and $\Phi_0= h/2e$.  The qubit frequency follows $\fqb = \sqrt{\Delta^2 + \varepsilon^2}$, where the tunnel coupling $\Delta = 2.6\GHz$ is fixed by fabrication and the energy detuning $\varepsilon = 2 I_\mathrm{P} \PhiQ/h$ is controlled by the applied flux $\Phi$ ($I_\mathrm{P}$ is the persistent current in the qubit loop).
The resonator frequency $\fr$ is around $2.3\GHz$ and depends only weakly on $\PhiQ$ and $\Isq$.  In addition, there are features visible at frequencies corresponding to the sum of the qubit and resonator frequencies, illustrating the coherent coupling between the two systems \cite{Chiorescu:2004, Goorden:2004}.

% Hamiltonian
The system is described by the Jaynes-Cummings Hamiltonian \cite{Blais:2004,BertetArxiv:2005,Bertet:2006}
\begin{equation}
  H/h = - \frac{\Delta}{2}\siX - \frac{\varepsilon}{2}\siZ  + \fr \left(a^\dag a + \frac{1}{2}\right) + \frac{g_1}{2}\left(a + a^\dag\right) \siZ.  \label{eq:H}
\end{equation}
Here, $a$ and $a^\dag$ are the creation/annihilation operators of the resonator field and $g_1$ is the dipole coupling between the qubit and the resonator.  In this work, we do not consider higher-order coupling parameters \cite{Bertet:2006,BertetArxiv:2005,Hoffman:2011}.

% - tunable qubit-LC circuit coupling
The coupling $g_1$ is mediated by the SQUID.  When the two SQUID junctions are symmetric, the current of the resonator mode splits equally into the two SQUID arms, and therefore no net flux is induced into the qubit loop.  The qubit is thus effectively decoupled from the resonator.
However, in the presence of a magnetic field, applying a dc bias current $\Isq$ creates an asymmetric phase drop over the two SQUID junctions.  This causes the resonator current to be slightly larger in one of the arms, which will produce a flux in the qubit loop.  The coupling to the resonator can thus be controlled \emph{in situ} by changing $\Isq$ \cite{Hime:2006}.

% flux induced in to the qubit loop
Figure \ref{fig:Sample}(c) shows the flux induced into the qubit loop as a function of the dc bias current $\Isq$ for the two devices.  Given the similarity in the design, both samples display similar behavior, with the bias current generating a parabolic shift in $\PhiQ$ \cite{Bertet:2005,Yoshihara:2006}.  Since $\PhiQ$ controls the qubit energy detuning $\varepsilon$, the first-order qubit-resonator coupling strength is determined by the derivative $\dd{\varepsilon}{\Isq} = (2 I_\mathrm{P}/h)(\dd{\PhiQ}{\Isq})$.  % (\dd{\varepsilon}{\PhiQ})(\dd{\PhiQ}{\Isq})
%
%\mpar{I still need to check the numbers here}
The bare coupling coefficient between qubit and resonator is then $g_1 = (\dd{\varepsilon}{\Isq}) \delta I_0$, where $\delta I_0 = \sqrt{2 \pi^2 h \fr^3 C_\mathrm{eff}}$ is the rms amplitude of the vacuum fluctuations and $C_\mathrm{eff} = 2C$ is the total capacitance of the resonant circuit \cite{Bertet:2006}.
%$\sqrt{h \fr/2L}$ $L$ inductance
% - coupling constants, in dE/dI_SQ and g1
The derivative $\dd{\varepsilon}{\Isq}$ and the coupling $g_1$ are plotted in \FigRef{fig:Sample}(d), determined at dc from the measured relation between $\PhiQ$ and $\Isq$ shown in \FigRef{fig:Sample}(c). Note that in both devices $g_1$ can be tuned over a range of a few tens of MHz, and that the coupling is turned off at $\Isq=\Isq^*$.  The device parameters are given in the figure caption.

\begin{figure}[tb]
\centering
\includegraphics[width=\linewidth]{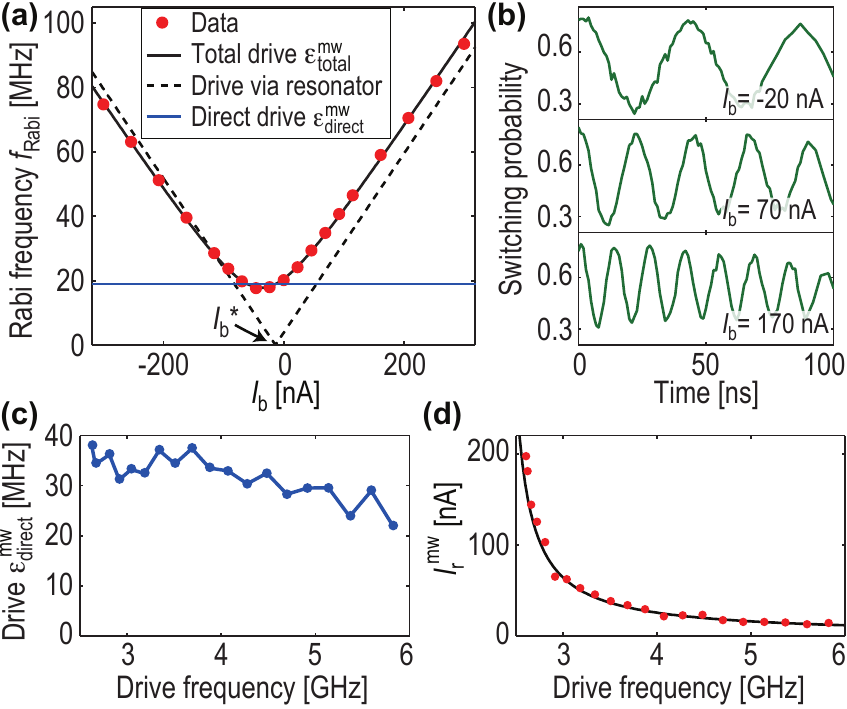}
\caption{(a) Rabi frequency of qubit A, measured vs $\Isq$ at $\PhiQ = 0$.  The driving field seen by the qubit contains two components: one is due to direct coupling to the antenna, the other is due to the coupling mediated by the resonator.
(b) Rabi traces for a few of the data points in panel (a).  The microwaves in the antenna have the same amplitude and frequency for all traces.
(c) Direct coupling between the antenna and the qubit, extracted from measurements similar to the one shown in panel (a).  The coupling depends only weakly on frequency.
(d) Microwave current in the resonator, induced by a fixed microwave amplitude in the antenna.  The black line is a fit to the square root of a Lorentzian, describing the oscillation amplitude of a harmonic oscillator with $f_r = 2.3\GHz$ and $Q = 100$.
} \label{fig:Rabi}
\end{figure}

% result device 1
Having determined the coupling coefficients, we turn to analyzing how the presence of the resonator influences the qubit's driven dynamics.
% - show fig2, describe methods
Figure \ref{fig:Rabi}(a) shows the extracted Rabi frequency $\fRabi$ of qubit A as a function of $\Isq$, measured at $\fqb=2.6\GHz$.  We find that $\fRabi$ changes by a factor of five over the range of the measurement, which is surprising since both the amplitude and the frequency of the microwave current $\Ian$ in the antenna are kept constant. The data points were obtained by fitting  Rabi oscillations to decaying sinusoids, a few examples of Rabi traces for different values of $\Isq$ are shown in \FigRef{fig:Rabi}(b).  We carefully calibrated the flux at each data point, to make sure that the qubit was driven on resonance and at the degeneracy point ($\varepsilon= 0$).

% - two drive methods
To explain the results of \FigRef{fig:Rabi}(a), we need to consider the indirect driving of the qubit mediated by the harmonic oscillator.  As seen in \FigRef{fig:Sample}(b), at zero flux detuning the qubit frequency ($\fqb = 2.6\GHz$) is relatively close to the resonator frequency ($\fr = 2.3\GHz$).   We therefore expect the microwave drive in the antenna to off-resonantly induce a microwave current $\Imr$ in the resonator, which is proportional to the square root of the average photon population, $\sqrt{\langle n \rangle}$.  By setting $\Isq \neq \Isq^*$, the coupling between the resonator and the qubit is turned on, and the resonator current $\Imr$ will start driving the qubit.
To describe this indirect driving, we treat the resonator classically and write the qubit Hamiltonian in \EqRef{eq:H} as
\begin{equation}
  H_\mathrm{qb}/h = - 1/2 \left[ \Delta \siX + \left[\varepsilon^\mathrm{dc} + \emt \cos(2\pi f t)\right] \siZ \right].
  \label{eq:H2}
\end{equation}
Here, the drive amplitude $\emt$ experienced by the qubit becomes a combination of  the drive $\emd$, due to direct coupling between antenna and qubit, and the drive $(\dd{\varepsilon}{\Isq}) \Imr$ mediated by the resonator.  We get: %g_1 \times \langle n \rangle =
\begin{equation}
 \emt  \!=\!
 \sqrt{\left[\emd\!+\!\cos \theta \ddf{\varepsilon}{\Isq} \Imr \!\right]^2
 \!+\! \left[\sin \theta \ddf{\varepsilon}{\Isq} \Imr\!\right]^2}\!,
   \label{eq:drive}
\end{equation}
where $\theta \equiv \theta_d-\theta_r $ is the phase difference between the direct drive and the drive mediated by the resonator. The Rabi frequency due to the drive $\emt$ depends on the qubit's quantization axis, which changes with the static energy detuning $\varepsilon_\mathrm{dc}$:
\begin{equation}
 f_\mathrm{Rabi} = \frac{\emt}{2} \frac{\Delta}{\sqrt{\varepsilon_\mathrm{dc}^2+\Delta^2}}.
   \label{eq:Rabi}
\end{equation}
%For a qubit driven resonantly, the measured Rabi frequency is $f_\mathrm{Rabi} = (\emt/2)\times (\Delta/\sqrt{\varepsilon^2+\Delta^2})$.
% - explain fit
Fitting the data in \FigRef{fig:Rabi}(a) to Eqs.~(\ref{eq:drive},\ref{eq:Rabi}) allows us to extract the parameters $\emd$, $I^\mathrm{mw}_\mathrm{r}$ and $\theta$. The different drive contributions are plotted together with the data in \FigRef{fig:Rabi}(a).  The direct drive is independent of $\Isq$, while the drive $\Imr (\dd{\varepsilon}{\Isq})$ mediated by the resonator increases linearly with $|\Isq|$, which originates from the linear dependence of $g_1$ shown in \FigRef{fig:Sample}(d). The minimum in Rabi frequency occurs at a value of $\Isq$ slightly shifted from the point $\Isq^*$ where $g_1=0$.  This offset appears because of the phase difference $\theta$ between the two drive components. The fit gives $\theta = -75^\circ$, which is consistent with a resonator driven above its resonance frequency.

% - go to higher qubit frequencies
Figures~\ref{fig:Rabi}(c) and \ref{fig:Rabi}(d) show how the two drive components depend on microwave frequency, measured by changing the static flux detuning $\PhiQ$ to increase the qubit frequency [see \FigRef{fig:Sample}(b)]. The direct drive only depends weakly on frequency (due to cable losses), whereas the drive mediated by the resonator drops sharply as the qubit-resonator detuning increases.  The black curve in \FigRef{fig:Rabi}(d) is the frequency response of a harmonic oscillator with $f_r = 2.3\GHz$ and $Q = 100$, with amplitude normalized to match the data.

\begin{figure}[t]
\centering
\includegraphics[width=\linewidth]{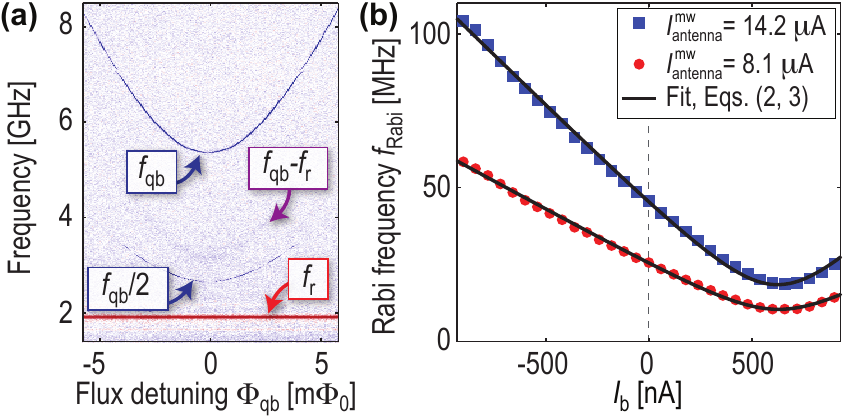}
\caption{(a) Spectrum of device B.  The spectral line at $2\GHz$ is the resonator, whereas the qubit tunnel coupling is $\Delta = 5.4\GHz$.
 (b) Rabi frequency vs bias current $\Isq$, measured at $f = 5.4\GHz$ and $\PhiQ = 0$ and for two different microwave drive amplitudes $\Ian$. Similar to device A, the Rabi frequency depends strongly on $\Isq$, and scales linearly with drive amplitude.  The black lines are fits to Eqs.~(\ref{eq:drive},\ref{eq:Rabi}), using the same coupling parameters for both sets of data.
 Note that the range of $\Isq$ in \FigRef{fig:RabiLongT1}(b) is several times larger than in \FigRef{fig:Rabi}(a).
} \label{fig:RabiLongT1}
\end{figure}

% - gap much larger than f_SQUID
To further investigate how the presence of the resonator affects the qubit dynamics at large detunings, we performed measurements on device B.  Figure \ref{fig:RabiLongT1}(a) shows a spectrum of that device, where the qubit and the resonator mode ($f_r = 2\GHz$) are clearly visible.
% - enable us to investigate influence on coherence
This device has a larger tunnel coupling ($\Delta = 5.4\GHz$), which allows us to operate the qubit at large frequency detuning from the resonator while still staying at $\varepsilon_\mathrm{dc} = 0$, where the qubit, to first order, is insensitive to flux noise \cite{Bertet:2005, Yoshihara:2006}. The qubit-resonator detuning corresponds to several hundred linewidths of the resonator, which is the regime of most interest for quantum information processing \cite{Majer:2007}.

% - show fig 3, similar results to fig 2, despite high off-resonant drive
Figure~\ref{fig:RabiLongT1}(b) shows the Rabi frequency vs bias current $\Isq$ of device B, measured at $f = 5.4\GHz$ and for two different values of the microwave drive current $\Ian$. Similarly to \FigRef{fig:Rabi}(a), the Rabi frequency clearly changes with $\Isq$, but the dependence is weaker than in \FigRef{fig:Rabi}(a) because of the larger frequency detuning.  Note that $\fRabi$ scales linearly with $\Ian$ for all values of $\Isq$.  By fitting the data to Eqs.~(\ref{eq:drive},\ref{eq:Rabi}), we find $\emd/\Ian = 6.4\MHz/\uA$, $I^\mathrm{mw}_\mathrm{r}/\Ian  = 2.4\nA/\uA$ and $\theta =-155^\circ$.
%We note that these numbers give a ratio $k = 0.15$ for $g_1=10\MHz$, which is close to the ratio for device A at $f = 5.4\GHz$ [see inset of \FigRef{fig:Rabi}(d)]. This is consistent with the similar geometries of the two samples. %$k = (2.4\nA/\uA)(0.4 \MHz/\nA)/(6.4\MHz/\uA)=0.15$
% - mention large phase shift
The large phase difference $\theta$ for device B causes the minimum in $\fRabi$ to shift away from the point close to $\Isq = 0$ where the coupling $g_1=0$ [see \FigRef{fig:Sample}(d)].
We attribute the large phase shift to influences from a second resonant mode, which is formed by the two $L$ and the two $C$ in the outer loop of \FigRef{fig:Sample}(a) \cite{Johansson:2006, Fedorov:2010}.  For sample B, this mode resonates around $5\GHz$.

% - issue with fluctuations in control parameter
The results of Figs.~\ref{fig:Rabi} and \ref{fig:RabiLongT1} show that the microwave signal mediated by the resonator plays a significant role when driving the qubit, appearing already at moderate qubit-resonator coupling $g_1$ and persisting even when the two systems are far detuned.
The design investigated here allows the coupling to be turned off [$g_1 = 0$ in \FigRef{fig:Sample}(d)], but it comes with a drawback: the parameter used to control the coupling ($\Isq$ in our setup) also provides a way for low-frequency noise to enter the system. Consider the relation between the $\fRabi$ and $\Isq$ in \FigRef{fig:RabiLongT1}(b): fluctuations $\delta \Isq$ near $\Isq = 0$ will cause fluctuations in the amplitude of the drive field seen by the qubit, which will lead to dephasing during driven evolution.

% rotary echo
% dephasing
To quantify the dephasing, we linearize the relation between the Rabi frequency and $\Isq$ close to $\Isq = 0$ as $f = f_0 [1+ r \, \delta I_b]$, where $f_0 = \fRabi(\Isq=0)$ and $r = (\dd{\fRabi}{\Isq})/f_0 = -1.28\,(\mu\mathrm{A})^{-1}$ is given by Eqs~(\ref{eq:drive},\ref{eq:Rabi}) or from \FigRef{fig:RabiLongT1}(b).
We model the fluctuations $\delta \Isq$ as normally distributed, with standard deviation $\sigma_I$.  Assuming the noise to be quasi-static, where the value of $\delta \Isq$ is constant during a single trial but differs from run to run \cite{Ithier:2005}, we find that the Rabi oscillations decay as
\begin{eqnarray}
   \int \frac{e^{-\delta \Isq ^2/(2\sigma_I^2)}}{\sqrt{2\pi\sigma_I^2}} \cos\left(2\pi f_0 [1+ r \, \delta \Isq ]  t\right) \, d\delta \Isq = \nonumber \\
     = e^{-2(\pi f_0 r \sigma_I t)^2} \cos\left(2\pi f_0 t\right).
   \label{eq:RabiDecay}
\end{eqnarray}
%
% - mention Rabi decay expected to be 4/3T1 + noise at S_omega (1/f)
In addition, the qubit energy relaxation time $T_1$ gives an exponential contribution to the Rabi decay, with time constant $4 T_1/3$ given by the Bloch equations.  The Rabi decay also depends on the flux noise at the Rabi frequency, but this contribution can be disregarded when operating the qubit at $\varepsilon_\mathrm{dc}=0$, where the qubit is insensitive to first-order flux noise \cite{Bylander:2011}.  The total decay envelope $f(t)$ of the Rabi oscillations becomes
\begin{equation}
 f(t) = e^{-\frac{3}{4T_1} t} e^{-(t/T_\varphi)^2},~\mathrm{with}~T_\varphi = 1/(\sqrt{2} \pi f_0 r \sigma_I).
   \label{eq:f}
\end{equation}

Note that the Gaussian decay constant $T_\varphi$ due to the effective amplitude fluctuations is inversely proportional to  $f_0$, the average Rabi frequency.  This is a consequence of having noise in the coupling between the qubit and the antenna; the effective amplitude fluctuations seen by the qubit will scale with the drive amplitude.
\begin{figure}[t]
\centering
\includegraphics[width=\linewidth]{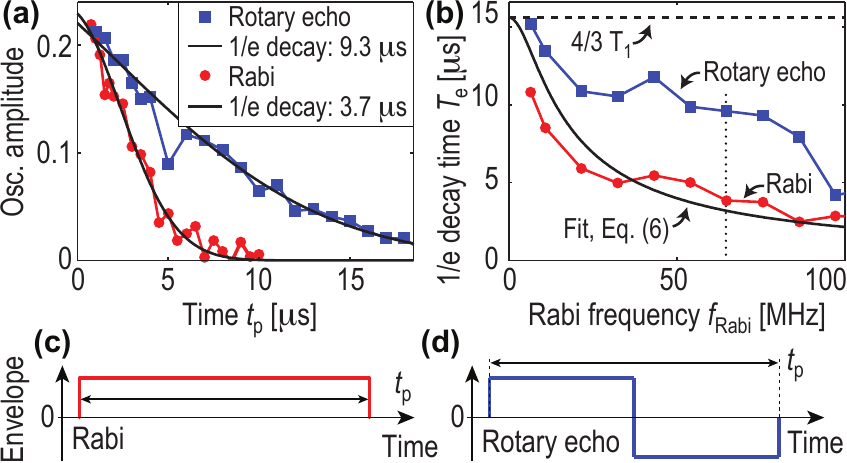}
\caption{(a) Decay envelopes of the Rabi and rotary-echo sequences for device B, measured with $\fRabi = 65\MHz$.  The solid lines are fits to \EqRef{eq:f}.
 (b) Decay times for Rabi and rotary echo, extracted from fits similar to the ones shown in panel (a).  The dashed line shows the upper limit set by qubit energy relaxation.  The dotted line marks the position for the decay envelope shown in panel (a).
 (c,d) Schematic diagrams describing the two pulse sequences in (a) and (b). For rotary echo, the phase of the microwaves is rotated by $180^\circ$ during the second half of the sequence.
} \label{fig:Rotary}
\end{figure}

The red circles in \FigRef{fig:Rotary}(a) show the envelope of Rabi oscillations measured for $\fRabi = 65\MHz$, together with a fit to \EqRef{eq:f}. The qubit energy relaxation $T_1=11.7\us$ is known from separate experiments \cite{Bylander:2011}, leaving $T_\varphi = 4.3 \us$ as a fitting parameter.
In \FigRef{fig:Rotary}(b), we plot the Rabi decay time versus $\fRabi$, extracted from envelopes similar to \FigRef{fig:Rotary}(a). To capture both the exponential and the Gaussian decay, we plot the time $T_e$ for the envelope to decrease by a factor $1/e$.  For the lowest Rabi frequency, the decay time is within $25\%$ of the upper limit set by qubit relaxation, but it decreases with $\fRabi$, as expected from \EqRef{eq:f}.  The black solid line shows a fit to \EqRef{eq:f}, giving a value of $\sigma_I=0.8\nA$ for the noise in $\Isq$.  The effective fluctuations in the drive amplitude are $r \sigma_I =0.06\%$.  We cannot rule out that part of that noise may be caused by instrument imperfections.

% - describe fix, rotary ech
Dephasing due to low-frequency fluctuations of the qubit frequency is routinely reduced by performing a Hahn-echo experiment \cite{Hahn:1950}.
Similarly, the fluctuations in drive field that cause decay of the Rabi oscillations in \FigRef{fig:Rotary}(b) can be mitigated with the rotary-echo pulse sequence \cite{Solomon:1959}, depicted in \FigRef{fig:Rotary}(c), which for driven systems is analogous to the Hahn-echo sequence. By shifting the phase of the drive by $180^\circ$ after a time $\tp/2$, any additional rotations, acquired due to slow fluctuations in the drive amplitude during the first half of the sequence, will cancel out during the reversed rotations in the second half of the sequence.

% - show trace in Fig 4b, describe measurement method
The blue squares in \FigRef{fig:Rotary}(a) shows the decay of the rotary-echo sequence, measured for $\fRabi = 65\MHz$.  The rotary-echo data shows a clear improvement compared to the Rabi decay for the same parameters [red circles in \FigRef{fig:Rotary}(a)].  We fit the rotary-echo data to \EqRef{eq:f} and plot the extracted decay times together with the results from Rabi measurements in \FigRef{fig:Rotary}(b).  The rotary-echo signal outperforms the Rabi decay over the full range of Rabi frequencies, and reaches the upper limit set by qubit relaxation $(4T_1/3)$ at low frequencies.  For intermediate frequencies, the rotary-echo decay times are slightly shorter than $4T_1/3$; we attribute the reduced coherence times to fluctuations in $\Isq$ that occur on time scales comparable to the length of the pulse sequence.  Noise at frequencies around $1/\tp$ will not be refocused by the reversed drive pulse, since the rotary-echo sequence has similar filtering properties as the Hahn-echo \cite{Solomon:1959}.   At the highest drive amplitudes ($\fRabi>100\MHz$), we observe a strong increase in decoherence, probably due to heating.
% two antennas
The indirect driving can also be reduced by driving the qubit with two antennas with different amplitudes and phases \cite{Groot:2010}, but it requires a more complicated setup.
% a setup with two separate microwave generators.

% conclusions
To conclude, we have investigated interference effects occurring when driving a qubit that is tunably coupled to a harmonic oscillator.  Although the addition of a coupling control parameter opens up an extra channel for dephasing, we show that its influence is reversible with dynamical decoupling techniques.  The results are relevant for any type of qubit that is tunably coupled to a resonator, and they show that despite engineering limitations, imperfections can be reversed by applying proper decoupling protocols.  In analogy with multi-pulse Hahn-echo experiments, we expect the incorporation of additional rotary echos to further improve the coherence times \cite{Bylander:2011}.

%\begin{acknowledgments}
We thank M. Gouker, X. Jin, and M. Neeley for helpful discussions and technical support. K.H. gratefully acknowledges the support of RIKEN and KFUPM (DSR FT100009).
%\end{acknowledgments}

%\bibliographystyle{apsrev}
\bibliographystyle{apsrev4-1}
\bibliography{IndirectDriving}

\end{document}